\PassOptionsToPackage{unicode}{hyperref}
\PassOptionsToPackage{hyphens}{url}
\PassOptionsToPackage{dvipsnames,svgnames,x11names}{xcolor}
\documentclass[
]{article}
\usepackage{amsmath,amssymb}
\usepackage{lmodern}
\usepackage{iftex}
\ifPDFTeX
  \usepackage[T1]{fontenc}
  \usepackage[utf8]{inputenc}
  \usepackage{textcomp} 
\else 
  \usepackage{unicode-math}
  \defaultfontfeatures{Scale=MatchLowercase}
  \defaultfontfeatures[\rmfamily]{Ligatures=TeX,Scale=1}
\fi
\IfFileExists{upquote.sty}{\usepackage{upquote}}{}
\IfFileExists{microtype.sty}{
  \usepackage[]{microtype}
  \UseMicrotypeSet[protrusion]{basicmath} 
}{}
\makeatletter
\@ifundefined{KOMAClassName}{
  \IfFileExists{parskip.sty}{%
    \usepackage{parskip}
  }{
    \setlength{\parindent}{0pt}
    \setlength{\parskip}{6pt plus 2pt minus 1pt}}
}{
  \KOMAoptions{parskip=half}}
\makeatother
\usepackage{xcolor}
\NewDocumentCommand\citeproctext{}{}
\NewDocumentCommand\citeproc{mm}{%
  \begingroup\def\citeproctext{#2}\cite{#1}\endgroup}
\makeatletter
 \let\@cite@ofmt\@firstofone
 \def\@biblabel#1{}
 \def\@cite#1#2{{#1\if@tempswa , #2\fi}}
\makeatother
\newlength{\cslhangindent}
\newlength{\csllabelwidth}
\newenvironment{CSLReferences}[2] 
 {\begin{list}{}{%
  \setlength{\itemindent}{0pt}
  \setlength{\leftmargin}{0pt}
  \setlength{\parsep}{0pt}
  \ifodd #1
   \setlength{\leftmargin}{\cslhangindent}
   \setlength{\itemindent}{-1\cslhangindent}
  \fi
  \setlength{\itemsep}{#2\baselineskip}}}
 {\end{list}}
\usepackage{calc}

\ifLuaTeX
\usepackage[bidi=basic]{babel}
\else
\usepackage[bidi=default]{babel}
\fi
\babelprovide[main,import]{american}

\def\languageshorthands#1{}
\ifLuaTeX
  \usepackage{selnolig}  
\fi
\IfFileExists{bookmark.sty}{\usepackage{bookmark}}{\usepackage{hyperref}}
\IfFileExists{xurl.sty}{\usepackage{xurl}}{} 
\hypersetup{
  pdftitle={Sequince: a modular sequential Monte Carlo toolkit},
  pdfauthor={Michael J. Williams, Christian E. A. Chapman-Bird},
  pdflang={en-US},
  colorlinks=true,
  linkcolor={Maroon},
  filecolor={Maroon},
  citecolor={Blue},
  urlcolor={Blue},
  pdfcreator={LaTeX via pandoc}}

\title{Sequince: a modular sequential Monte Carlo toolkit}

\definecolor{c53baa1}{RGB}{83,186,161}
\definecolor{c202826}{RGB}{32,40,38}

\usepackage[affil-it]{authblk}
\usepackage{orcidlink}
\author[1,2%
  ]{Michael J. Williams%
    \,\orcidlink{0000-0003-2198-2974}\,%
    }
\author[3%
  ]{Christian E. A. Chapman-Bird%
    \,\orcidlink{0000-0002-2728-9612}\,%
    }

\affil[1]{Max Planck Institute for Gravitational Physics (Albert
Einstein Institute), Am Mühlenberg 1, 14476 Potsdam, Germany%
  }
\affil[2]{Institute of Cosmology and Gravitation, University of
Portsmouth, United Kingdom%
  }
\affil[3]{Institute for Gravitational Wave Astronomy, University of
Birmingham, United Kingdom%
  }
\date{\today}

\begin{document}
\maketitle
\begin{abstract}
Sequential Monte Carlo (SMC) methods approximate sequences of
probability distributions using populations of weighted samples and are
widely applicable, including in Bayesian inference and particle
filtering. We present \texttt{sequince}, a modular Python toolkit for
constructing SMC algorithms from reusable and extensible components. The
toolkit provides modular steps for weighting, resampling, mutation,
transformation, and control operations that are composed to implement
different algorithms. It also includes prebuilt samplers for tempered
SMC, persistent sampling and nested sampling via SMC. Its core
components follow the Python Array API standard, supporting NumPy, JAX,
PyTorch, and CuPy, enabling analyses on CPUs or GPUs. Random-number
control and checkpointing support reproducible scientific workflows.
This paper describes \texttt{sequince} v0.1.0 and its design as reusable
infrastructure for methodological research and application of SMC
methods.
\end{abstract}

\section{Summary}\label{summary}

Sequential Monte Carlo (SMC) methods are a flexible family of algorithms
for approximating sequences of probability distributions using a
population of weighted samples, known as particles
(\citeproc{ref-chopin2020introduction}{Chopin \& Papaspiliopoulos,
2020}; \citeproc{ref-delmoral2006smc}{Del Moral et al., 2006}). Such
sequences arise naturally in applications including Bayesian inference
and particle filtering, for example through likelihood tempering or
observations arriving over time.

\texttt{sequince} (\citeproc{ref-sequince}{Williams \& Chapman-Bird,
2026}) is a Python toolkit for constructing SMC methods from reusable
and extensible components. This paper describes \texttt{sequince}
v0.1.0. The toolkit provides abstractions for defining sequences of
target distributions and for composing the operations used to evolve a
particle population between them. These include weighting, resampling,
mutation, and control steps. It is designed for researchers developing
new SMC methods or adapting existing methods to scientific applications,
including astrophysical inference, with checkpointing to support
long-running analyses. Its core components follow the Python Array API
standard, supporting NumPy, JAX, PyTorch, and CuPy arrays and enabling
compatible workflows to run on CPUs or GPUs.

\section{Statement of need}\label{statement-of-need}

Scientific applications often require SMC methods tailored to a
particular problem. Researchers may need to combine adaptive tempering,
mutation kernels, transformations, stopping rules, or multi-stage
procedures, and to define problem-specific sequences of target
distributions. When this logic is embedded in application-specific code,
it becomes harder to compare methods, reuse components, and move
prototype algorithms into stable analysis software.

\texttt{sequince} addresses these needs through a general-purpose,
step-oriented framework with interchangeable algorithmic components.
Researchers can start from pre-built samplers for common inference
problems or construct custom pipelines with explicit control over the
operations and their execution order. The same framework therefore
supports both established inference methods and methodological research
into new SMC variants.

Scientific models are also increasingly implemented using different
numerical libraries to exploit automatic differentiation and hardware
acceleration. Coupling an SMC implementation to a single array library
can restrict its integration with these models. \texttt{sequince} uses
the Python Array API standard
(\citeproc{ref-python_array_api}{Consortium for Python Data API
Standards, 2026}) to support reusable core operations across multiple
numerical backends, while allowing specialised components to target
particular ecosystems.

\section{State of the field}\label{state-of-the-field}

The Python ecosystem already contains several useful SMC-related tools.
The \texttt{particles} package accompanying Chopin \& Papaspiliopoulos
(\citeproc{ref-chopin2020introduction}{2020}) provides implementations
of many established SMC and particle-filtering methods and is
particularly useful for teaching and methodological prototyping.
\texttt{SMCPy} focuses on uncertainty quantification and Bayesian
calibration using parallel likelihood-tempering SMC samplers, including
fixed and adaptive tempering variants (\citeproc{ref-smcpy_repo}{NASA
Langley Research Center, 2026}). \texttt{pocomc} implements persistent
sampling with normalizing flows for posterior inference
(\citeproc{ref-karamanis2025persistent}{Karamanis \& Seljak, 2025}).

Broader Bayesian software ecosystems also expose SMC functionality.
\texttt{PyMC} includes a sequential Monte Carlo sampler within a general
probabilistic programming framework
(\citeproc{ref-abrilpla2023pymc}{Abril-Pla et al., 2023}), while
\texttt{BlackJAX} provides composable inference components in the JAX
ecosystem and includes SMC methods alongside other Monte Carlo
algorithms (\citeproc{ref-cabezas2024blackjax}{Cabezas et al., 2024}).
Recent domain-specific frameworks such as SHARPy target
gravitational-wave inference with JAX-accelerated SMC and gradient-based
mutation kernels (\citeproc{ref-demasi2026sharpy}{Demasi et al., 2026}).

\texttt{sequince} combines support for multiple SMC variants, including
tempering, persistent sampling
(\citeproc{ref-karamanis2025persistent}{Karamanis \& Seljak, 2025}),
nested sampling via SMC (NS-SMC)
(\citeproc{ref-salomone2025unbiased}{Salomone et al., 2025}), and
waste-free SMC (\citeproc{ref-dau2022wastefree}{Dau \& Chopin, 2022}),
with a modular architecture for constructing bespoke algorithms. Shared
interfaces allow users to reconfigure target sequences, transformations,
mutation kernels, and control flow, supporting both the comparison of
existing methods and the development of new variants. Support for
multiple array backends and integrations with different sampling
ecosystems further enable these components to be reused across
scientific applications and numerical libraries. The need to combine
configurable target sequences, independently reusable steps, and
multiple array backends within one framework motivated this design. This
combination makes \texttt{sequince} useful as research infrastructure
for methodological work and as a foundation for application-specific
downstream packages.

\section{Software design}\label{software-design}

\texttt{sequince} separates target distributions, particle populations,
and the operations used to evolve them. The following outlines the main
design choices; the
\href{https://sequince.readthedocs.io/en/latest/}{documentation} and
\href{https://sequince.readthedocs.io/projects/sequince-examples/en/latest/}{runnable
examples} provide detailed descriptions of the interfaces and their use.

The distributions approximated during an SMC run are represented by
target objects. A target evaluates named components of a probability
distribution for a population of particles, while sequential targets
describe how that distribution changes between stages. Separating the
target from the operations that evolve the particles allows weighting
and mutation components to be reused across different inference
procedures. For example, tempered SMC and nested-sampling SMC use
different target sequences while sharing resampling and compatible
mutation components.

Target evaluations are stored as particle-aligned collections of named
quantities. These can be propagated through resampling and updated
during mutation, allowing previously computed quantities to be reused
where valid. This is particularly useful for scientific applications in
which likelihood evaluations dominate the computational cost. A
\texttt{ParticleState} stores the particles, their log weights,
accumulated normalising-constant information, and associated target
evaluations. Specialised states provide additional information where
required, such as the archive of previous populations used by persistent
SMC.

An SMC algorithm is expressed as an ordered collection of steps. Each
step receives the current particle state, a runtime context, and a
random number generator, and returns an updated state, updated context,
and structured outputs. Weighting, resampling, mutation, transformation,
and control operations can therefore be implemented independently and
combined into different algorithms. Higher-level steps can conditionally
execute or compose other steps, keeping algorithm-specific control flow
outside the runner. For example, waste-free SMC combines resampling with
a mutation component that retains intermediate chain states.

A \texttt{SequentialMonteCarlo} runner repeatedly applies the configured
steps. The runtime context contains the current iteration and
observation, shared user-provided information, metadata used for
communication between steps, and stopping controls. The runner accepts
an iterable of observations, supporting both static inference problems
and procedures in which targets change as new data become available.
Separating particle state, runtime context, and algorithm configuration
allows the execution machinery to remain common across different SMC
constructions.

Each step can return structured outputs containing quantities such as
effective sample sizes, tempering parameters, acceptance statistics, and
other diagnostics. These outputs and the recorded particle states are
collected in an \texttt{SMCResult} for subsequent analysis. New
components can therefore introduce diagnostics without requiring changes
to the runner or coupling their implementation to plotting and logging
code. Callbacks provide logging, plotting, and other monitoring
functionality during execution.

The core abstractions follow the Python Array API standard where
possible (\citeproc{ref-python_array_api}{Consortium for Python Data API
Standards, 2026}), allowing backend-native arrays to pass through the
SMC procedure. Random number generation is similarly abstracted through
\texttt{orng} (\citeproc{ref-orng}{Williams \& Talbot, 2026}). This
design supports reuse across numerical libraries while allowing
specialised components to depend on a particular ecosystem. Alongside
conventional eager execution, \texttt{sequince} provides a functional
execution path that passes random state explicitly between steps.
Compatible iterations can be compiled using facilities such as
\texttt{jax.jit}, provided their operations are traceable and their
state structures, shapes, and dtypes remain stable, while the runner
manages observations, callbacks, and result recording. Mutation
integrations include ecosystem-specific packages such as
\texttt{BlackJAX} (\citeproc{ref-cabezas2024blackjax}{Cabezas et al.,
2024}).

Checkpointing is separate from result recording. Checkpoints capture the
particle state, runtime progress, random-number state, and mutable step
state needed to continue sampling with a reconstructed sampler
configuration. Results provide the analysis record of a completed or
partially completed run. Both support HDF5 storage, making numerical
data and supported metadata available for inspection independently of
the running process.

\section{Research impact statement}\label{research-impact-statement}

\texttt{sequince} is being developed as reusable infrastructure for both
methodological research and problem-specific inference workflows.
Current applications include customised sampling strategies for
astrophysical inference problems relevant to ongoing LIGO-Virgo-KAGRA
analyses and future observations with the Laser Interferometer Space
Antenna (LISA) (\citeproc{ref-williams2026inprep}{Williams et al.,
2026}), as well as integration with broader Bayesian inference
ecosystems such as \texttt{bilby} (\citeproc{ref-ashton2019bilby}{Ashton
et al., 2019}; \citeproc{ref-bilby}{Talbot et al., 2026}) via
\texttt{sequince-bilby} (\citeproc{ref-sequince_bilby}{Williams,
2026b}). The same methodology is also relevant to future ground-based
observatories such as the Einstein Telescope (ET) and Cosmic Explorer
(CE).

In these applications, the principal benefit of \texttt{sequince} is not
a single sampling algorithm but a shared framework in which new
sequential inference methods can be implemented, compared, and reused.
The package has been used to explore alternative tempering strategies,
persistent SMC, specialised mutation schemes, problem-specific
transformations, and hardware-accelerated execution. The common result
interface also provides a consistent way to record and compare
algorithm-specific diagnostics across these different methods.

The modular design of \texttt{sequince} additionally supports downstream
packages that add method- or domain-specific functionality without
requiring those dependencies in the core library. For example,
\texttt{sequince-density} (\citeproc{ref-sequince_density}{Williams,
2026c}) provides density-estimation components based on Gaussian mixture
models and normalizing flows. These can be combined with the core SMC
machinery for learned preconditioning, proposal construction, and other
density-assisted SMC methods, including approaches related to
\texttt{pocomc} and \texttt{aspire}
(\citeproc{ref-karamanis2025persistent}{Karamanis \& Seljak, 2025};
\citeproc{ref-williams2026accelerated}{Williams, 2026a}). The wider
sequince ecosystem also includes \texttt{sequince-examples}, which
provides runnable examples demonstrating common workflows and more
specialised uses of the package
(\citeproc{ref-sequince_examples}{Sequince contributors, 2026}), with
\href{https://sequince.readthedocs.io/projects/sequince-examples/en/latest/}{online
documentation}.

\section{AI usage disclosure}\label{ai-usage-disclosure}

Generative AI tools, including OpenAI Codex, were used to assist with
code implementation and refactoring. ChatGPT was also used to assist
with the organisation and editing of the manuscript. Models varied over
the course of development and included GPT-5.3, GPT-5.4, GPT-5.6,
GPT-6.0, and GPT-6.1; a complete historical record of model versions was
not retained. The authors made the core scientific and software design
decisions and reviewed, edited, and validated all AI-assisted
contributions. The authors retain full responsibility for the software
and manuscript.

\section{Acknowledgements}\label{acknowledgements}

M.J.W. acknowledges partial support from the UKSA's LISA Ground Segment
(UKRI968) and Science and Technology Facilities Council (STFC) grant
UKRI2490. C.E.A.C-B. is supported by UKSA grant UKRI971.

Numerical computations were carried out on the SCIAMA high-performance
computing cluster, which is supported by the Institute of Cosmology and
Gravitation and the University of Portsmouth. Additional computations
were performed using the University of Birmingham's BlueBEAR High
Performance Computing facility.

\section*{References}\label{references}
\addcontentsline{toc}{section}{References}

\protect\phantomsection\label{refs}

\end{document}